\documentclass{article}

\usepackage{microtype}
\usepackage{subcaption}
\usepackage{booktabs} 

\usepackage{graphicx}   
\usepackage{setspace}   
\usepackage{multirow}   
\usepackage[table]{xcolor} 

\usepackage{hyperref}

\newcommand{\system}{\textsc{IntOpt}\xspace} 

\usepackage[preprint]{icml2026}

\usepackage{amsmath}
\usepackage{amssymb}
\usepackage{mathtools}
\usepackage{amsthm}
\usepackage{xspace}
\usepackage{listings}
\usepackage{graphicx}
\usepackage{adjustbox}
\usepackage{tcolorbox}
\usepackage{setspace}
\usepackage{multirow}
\usepackage{makecell}
\usepackage{booktabs}
\usepackage{enumerate}
\usepackage{enumitem}
\usepackage[capitalize,noabbrev]{cleveref}

\theoremstyle{plain}

\theoremstyle{definition}

\theoremstyle{remark}

\newcommand{\hl}[1]{\textcolor{green!70!black}{\textbf{#1}}}   
\newcommand{\hltext}[1]{\textcolor{green!70!black}{#1}}        

\lstdefinestyle{mylisting}{
  frame=single,
  rulecolor=\color{white},
  basicstyle=\ttfamily\small,
  breaklines=true,
  breakautoindent=false,
  showstringspaces=false,
  escapeinside={(*@}{@*)}
}

\usepackage[textsize=tiny]{todonotes}

\icmltitlerunning{Submission and Formatting Instructions for ICML 2026}

\begin{document}

\twocolumn[
    \icmltitle{Beyond Pass-by-Pass Optimization: \\
    Intent-Driven IR Optimization with Large Language Models}




  \begin{icmlauthorlist}
    \icmlauthor{Lei Qiu}{ict,ucas}
    \icmlauthor{Zi Yang}{ict,jnu}
    \icmlauthor{Fang Lyu}{ict}
    \icmlauthor{Ming Zhong}{cuhk}
    \icmlauthor{Huimin Cui}{ict,ucas}
    \icmlauthor{Xiaobing Feng}{ict,ucas}
  \end{icmlauthorlist}

  \icmlaffiliation{ict}{SKLP, ICT, CAS, China}
  \icmlaffiliation{ucas}{UCAS, China}
  \icmlaffiliation{jnu}{Jiangnan University, China}
  \icmlaffiliation{cuhk}{CUHK, China}

  \icmlcorrespondingauthor{Fang Lyu}{flv@ict.ac.cn}
  \icmlcorrespondingauthor{Huimin Cui}{cuihm@ict.ac.cn}

  \icmlkeywords{Machine Learning, ICML}

  \vskip 0.3in
]



\printAffiliationsAndNotice{}  

\begin{abstract}
Modern compilers optimize programs through a sequence of modular passes over intermediate representations (IR). While this pass-by-pass paradigm offers engineering benefits, it suffers from a pass coordination problem: locally beneficial transformations may block more profitable optimizations in later stages. This limitation stems from the lack of an explicit notion of optimization intent, defined as a holistic strategy for coordinating multiple transformations toward a global performance objective.
Recent LLM–based approaches formulate IR optimization as an end-to-end generation task, thereby avoiding the traditional pass-by-pass structure. However, optimization intent remains implicit in these methods, forcing models to jointly infer optimization strategy and generate low-level transformations, which limits both correctness and performance.
We propose IntOpt, the first intent-driven IR optimizer that explicitly separates high-level optimization intent from low-level analysis and transformation. IntOpt organizes IR optimization into three stages: intent formulation, intent refinement, and intent realization, enabling globally coordinated transformations. 
Experiments show that IntOpt achieves 90.5\% verified correctness and 2.660× average speedup on 200-program test set, outperforming state-of-the-art LLM-based optimizers in both correctness and performance, and surpassing modern compiler with the \texttt{-O3} option on 37 benchmarks with speedups of up to 272.60×.
\end{abstract}
\section{Introduction}\label{sec:intro}

Modern compilers leverage intermediate representations (IR) to enable compilation that is independent of both programming languages and hardware architectures~\cite{dragon_book, LLVM_paper}. By optimizing at the IR level, which abstracts away language- and hardware-specific details, compilers can improve the program execution efficiency (referred to as ``\textbf{performance}") across diverse programming languages, target architectures, and application domains, ranging from traditional scientific computing~\cite{10.1145/3572848.3577475, Julia, 8639434} to modern AI workloads with specialized tensor operations~\cite{mlir, tvm, xla}. Consequently, designing effective IR optimization strategies has remained significant in compiler construction.

However, the effectiveness of IR-level optimization is fundamentally constrained by \emph{how modern compilers organize optimization}. In practice, modern compilers decompose IR optimization into a sequence of passes, each implemented as a modular unit that targets a specific optimization, such as loop unrolling or constant propagation~\cite{LLVM_paper, ir_survey}. These passes make decisions by analyzing optimization-related information derived from the IR (hereafter referred to as \textbf{``analysis''}) to assess whether a transformation will be beneficial. Although this \textbf{pass-by-pass paradigm} offers clear engineering benefits in terms of modularity and maintainability, it inherently suffers from a pass coordination problem: optimizations that are locally beneficial in early passes can inadvertently restrict or even prevent more profitable optimizations in later stages.

\begin{figure*}
    \centering
    \includegraphics[width=0.85\linewidth]{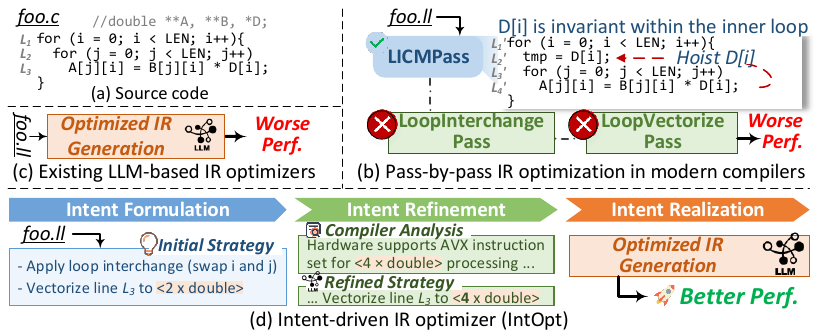}
    \vspace{-1.5ex}
    \caption{Comparison between existing IR optimization approaches and our \system. }
    \label{fig:cmp}
    \vspace{-2.5ex}
\end{figure*}

As illustrated in \autoref{fig:cmp}(b), when the compiler optimizes the IR corresponding to the source code in \autoref{fig:cmp}(a), the \textit{LICMPass} (loop-invariant code motion) hoists \texttt{D[i]} from $L_3$ to $L_2'$ to eliminate redundant accesses in the inner loop. While this transformation is locally beneficial, it prevents the \textit{LoopInterchangePass} from swapping the \texttt{i} and \texttt{j} loops to improve memory locality. More critically, because \textit{LoopInterchangePass} cannot be applied, the subsequent \textit{LoopVectorizePass} also fails to vectorize $L_4'$. As a result, the innermost loop retains a column-major access pattern for \texttt{A[j][i]} and \texttt{B[j][i]}, which inhibits effective SIMD utilization and ultimately leads to degraded performance.

This behavior arises from the lack of an explicit notion of \textbf{optimization intent}, which we define as a holistic strategy for coordinating multiple transformations toward a global performance objective. In compilers such as GCC~\cite{Web:GCC} and LLVM~\cite{Web:LLVM}, this intent is implicit and fragmented across individual optimization passes. As shown in \autoref{fig:cmp}(b), \textit{LICMPass} reduces memory operations, \textit{LoopInterchangePass} improves memory locality, and \textit{LoopVectorizePass} targets SIMD optimization. However, without explicit global coordination, individual passes cannot anticipate how their transformations may enable or inhibit downstream optimization opportunities~\cite{1611550, 8130281}.

Recent advances in large language models (LLMs)~\cite{Web:gpt5, Web:claude, deepseek-nature} offer a promising way to address this limitation due to their strong program understanding and semantic reasoning capabilities. Prior LLM-based IR optimization methods~\cite{cummins2025llm, yang2025iroptset} formulate IR optimization as an end-to-end generation task that directly maps unoptimized IR to optimized IR, as shown in \autoref{fig:cmp}(c), avoiding the traditional pass-by-pass structure in compilers. However, this reformulation does not fundamentally resolve the core issue: optimization intent remains implicit, now hidden within a black-box LLM generation process rather than fragmented across individual passes in compilers. Consequently, LLMs are forced to simultaneously infer optimization intent and generate precise transformations, which makes it challenging to guarantee both semantic correctness and consistent performance improvement. Our experiments in \autoref{sec:imp_llmOpt} empirically reveal substantial room for improvement.

In this paper, we propose \system, an \underline{Int}ent-driven IR \underline{Opt}imizer that reformulates the modern compiler optimization paradigm by decoupling high-level optimization intent from low-level analysis and transformation, thereby enabling globally coordinated transformations. \system achieves this goal by deconstructing the IR optimization process into three stages: \textbf{intent formulation}, \textbf{intent refinement}, and \textbf{intent realization}. This pipeline mirrors how expert compiler engineers operate: first reasoning about optimization intent at a global level, then grounding it using program analysis to assess feasibility, and finally realizing it through concrete transformations~\cite{10.5555/286076}.

Our key insight is that, although existing compilers encode optimization intent implicitly within hand-crafted passes, these intents capture valuable domain knowledge that LLMs can learn and make explicit. Building on this, \system first uses an LLM specialized in compiler optimization to infer an initial optimization strategy for a given program, representing a structured form of optimization intent. This strategy is then grounded and refined using compiler analysis. For example, in \autoref{fig:cmp}(d), an initial strategy of vectorization with \texttt{<2 x double>} is refined to \texttt{<4 x double>} based on hardware-aware analysis. The refined strategy subsequently guides optimized IR generation, ensuring that transformations are globally coordinated. During the intent refinement and realization stage, \system leverages GPT-5~\cite{Web:gpt5} and its pre-training knowledge to explore optimization opportunities beyond those captured by existing compiler passes, as shown in \autoref{sec:case_study}.

We evaluate \system on 200 LLVM IR programs and achieve significant improvements over both end-to-end LLM-based optimizers and traditional compiler. Our results highlight the importance of making optimization intent explicit for globally coordinated and semantically grounded IR optimization. We hope our findings will facilitate more research on intent-driven compiler optimization in this field.

\begin{figure*}[th]
    \centering
    \includegraphics[width=0.95\linewidth]{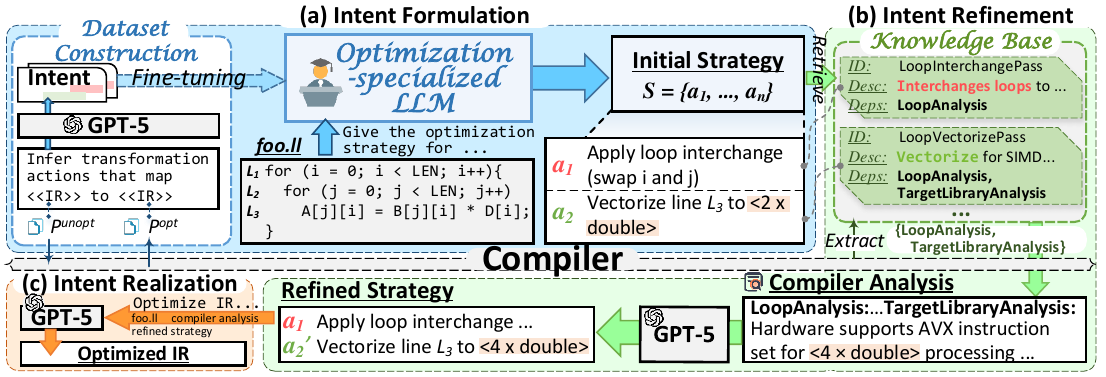}
    \caption{Overview of \system.}
    \label{fig:overview}
    \vspace{-2ex}
\end{figure*}

To sum up, our main contribution includes:
\vspace{-1.5ex}
\begin{itemize}[itemsep=1pt, leftmargin=2ex]
    \item \textbf{Intent-driven IR optimization framework.} We propose the first intent-driven IR optimization framework that decouples optimization intent from concrete IR transformation, enabling more effective and reliable optimizations.

    \item \textbf{Compiler-knowledge-grounded IR optimization.} We develop a novel approach that integrates structured compiler knowledge to guide LLMs in IR optimization.
    
    \item \textbf{Comprehensive evaluation.} \system achieves 90.5\% verified correctness and 2.660$\times$ average speedup on 200-program test set, outperforming state-of-the-art LLM-based optimizers by 7.0\%-31.5\% in correctness and up to 1.603$\times$ in performance, while also surpassing LLVM \texttt{-O3} on 37 benchmark programs with speedups up to 272.60$\times$.
    
\end{itemize}
\section{Related Works}

\textbf{Traditional IR Optimization in Compilers.}
Modern compilers such as LLVM~\cite{Web:LLVM} and GCC~\cite{Web:GCC} perform IR optimization through a fixed sequence of passes (e.g., \texttt{-O3}), where each pass applies a specific transformation guided by a predefined set of compiler analysis and hand-crafted heuristics~\cite{LLVM_paper}. To provide flexibility, compilers expose numerous configuration options that allow users to enable or disable specific optimizations or adjust heuristic parameters~\cite{10.5555/286076}. However, interactions among optimization passes are often complex and difficult to predict~\cite{1611550, 8130281}, making it challenging to reason about or control global optimization strategies for a given program.

\textbf{Autotuning for Compiler Optimization.}
Building on the pass-by-pass optimization infrastructure, a line of work~\cite{NVP, opentuner, BOCA, CompTuner, Stuner, 10.1145/3696443.3708961} has explored autotuning approaches that search over compiler options, optimization flags, or pass configurations to improve performance. 
While these methods can yield performance gains, it suffers from several inherent limitations. First, the search process is time-consuming, as it requires repeated compilation (and, for runtime performance–oriented objectives, execution) under many configurations. Second, optimization options are coarse-grained: disabling, enabling, or reordering a pass affects all occurrences of that transformation across the entire program. For example, if loop-invariant code motion is disabled at the option level to avoid harming a particular loop (as in \autoref{fig:cmp}(a)), other loops where LICM would be beneficial are also affected. 

\textbf{LLMs for Compiler and Code Optimization.}
Recent work has begun to explore using LLMs for compiler construction and program optimization. Some approaches focus on enhancing compiler backend development efficiency~\cite{zhong2024comback,zhong2025bepilot,zhong2025VEGA}, while others target source-level program rewriting for vectorization~\cite{LLM-Vec}. More closely related to our work are efforts that apply LLMs directly to IR-level optimization~\cite{cummins2025llm, yang2025iroptset}, formulating optimization as an end-to-end generation task. While these methods bypass the traditional pass-by-pass pipeline, they typically rely on black-box generation without explicit representations of optimization intent. 
In contrast, our work models optimization intent as an explicit intermediate step and integrates LLM-based reasoning with compiler knowledge to enable globally coordinated IR optimization.

\section{Proposed Method: \system}

To explicitly model and reason about optimization intent during compiler optimization, we propose \system, an intent-driven IR optimization framework. As illustrated in \autoref{fig:overview}, \system contains three stages: intent formulation (\autoref{sec:intent_formulation}), intent refinement (\autoref{sec:intent_refinement}), and intent realization (\autoref{sec:intent_realization}). This structure decouples high-level optimization intent from low-level analysis and transformations, allowing \system to make globally coordinated IR optimization.

\subsection{Intent Formulation}
\label{sec:intent_formulation}

The intent formulation stage generates an initial optimization strategy $S$, which serves as a structured instantiation of the optimization intent, for a given input IR program $P$:
$$S = \mathcal{F}_{\theta}(P)$$
where $\mathcal{F}_{\theta}$ denotes an optimization-specialized LLM parameterized by $\theta$, fine-tuned on compiler optimization behaviors. The strategy $S$ is represented as an ordered sequence of transformation actions $\langle a_1, a_2, \ldots, a_n \rangle$, where each action $a_i$ specifies in natural language which transformation to apply to which program region (e.g., ``Vectorize line $L_3$ with \texttt{<2 x double>}").

Representing actions in natural language avoids prematurely constraining the optimization action space, allowing the model to reason over a richer and more flexible space of optimization strategies. Despite this flexibility, the problem remains tractable: while the space of semantically equivalent IR programs is vast, the set of transformation actions exposed by a modern compiler is finite and curated by compiler developers. This bounded action set makes intent formulation amenable to data-driven learning from a modestly sized dataset. Consequently, to fine-tune the intent formulation model $\mathcal{F}_{\theta}$, we construct a dataset by distilling optimization intents from compiler behaviors.

\textbf{Dataset Construction.}
As shown in \autoref{fig:overview}(a), we first collect a diverse set of IR programs from IR-OptSet~\cite{yang2025iroptset}, 
an IR-oriented dataset designed to trigger representative optimization transformations encountered in real compiler optimization scenarios.
We compile the collected programs using LLVM 19.1.0~\cite{Web:LLVM} with the \texttt{-O3} flag to obtain paired unoptimized and optimized IR programs $(P^{\text{unopt}}, P^{\text{opt}})$, where \texttt{-O3} enables a broad range of optimization passes.
We then distill optimization intent from the observed end-to-end optimization behavior $P^{\text{unopt}}$ to $P^{\text{opt}}$ by using GPT-5~\cite{Web:gpt5} to infer the underlying optimization strategy $S$ for each IR pair. 
The resulting triples -- unoptimized IR $P^{\text{unopt}}$, optimized IR $P^{\text{opt}}$, and inferred strategy $S$ -- constitute the training dataset $\mathcal{D}_{\text{train}}$.

\textbf{Model Training.} 
We fine-tune LLM Compiler FTD 13B~\cite{cummins2025llm}, an LLM for IR optimization, on $\mathcal{D}_{\text{train}}$ to obtain the optimization-specialized LLM $\mathcal{F}_{\theta}$. The training objective is to minimize the negative log-likelihood of generating the ground-truth optimization strategy $S$ conditioned on the unoptimized IR program $P^{\text{unopt}}$:
$$
\mathcal{L}(\theta)
= - \mathbb{E}_{(P^{\text{unopt}}, S) \sim \mathcal{D}_{\text{train}}}
\left[ \log p_{\theta}(S \mid P^{\text{unopt}}) \right]
$$
The choice of a 13B-parameter model is deliberate: intent formulation is primarily pattern recognition over program structures and compiler optimization behaviors, which does not require the extensive reasoning capabilities needed for subsequent refinement and realization stages. This design balances effectiveness with computational efficiency.

\subsection{Intent Refinement}
\label{sec:intent_refinement}

The intent formulation stage produces an initial optimization strategy that specifies which transformations to apply and in what order.
Although this strategy captures high-level optimization intent learned from compiler behaviors, it is inferred in a data-driven manner and may miss key constraints from program semantics or target-specific requirements. The intent refinement stage therefore grounds the initial strategy in concrete compiler analysis and revises it when needed, ensuring that each proposed transformation is applicable and well-informed.

A key challenge of intent refinement is identifying which compiler analysis is necessary to refine the given strategy $S$. Since different transformations rely on different types of analysis, retrieving all available analysis is both inefficient and unnecessary.

To address this challenge, we design an analysis-aware retrieval mechanism that explicitly captures the relationship between transform passes and their required analysis. By constructing a structured compiler knowledge base and using it to retrieve only the analysis relevant to each transformation action in $S$, the refinement process concentrates on the most informative analysis signals while avoiding unnecessary overhead.

\textbf{Compiler Knowledge Base Construction.} We construct a structured knowledge base $\mathcal{K}$ that catalogs compiler optimization passes (i.e., transform passes) and their analysis dependencies. Each entry $k \in \mathcal{K}$ is represented as:
$$
k = (ID, Desc, Deps),
$$
where $ID$ uniquely identifies a transform pass (corresponding to its pass class name, e.g., \texttt{LoopVectorizePass}), $Desc$ provides a natural-language description of the pass’s functionality and optimization objectives, and $Deps$ denotes the set of compiler analysis required by the pass.

We build $\mathcal{K}$ through an automatic extraction process from the LLVM compiler infrastructure. First, we parse \texttt{llvm/lib/Passes/PassRegistry.def}, which enumerates all transform passes, to obtain the set of pass identifiers $ID$. For each pass, we then extract its functional description from the official LLVM pass documentation\footnote{\url{https://llvm.org/docs/Passes.html}} to populate the corresponding $Desc$ field.

To extract the analysis dependencies $Deps$, we exploit a key design principle of LLVM: each transform pass must explicitly declare analysis passes it relies on by invoking \texttt{AnalysisManager::getResult} within its \texttt{run} method. Accordingly, LLVM provides a collection of analysis passes, each responsible for extracting specific optimization-related information from the IR programs to guide transformation decisions. This explicit dependency mechanism allows us to systematically trace which analysis passes are required by each transform pass, yielding a reliable and complete specification of its analysis requirements. For example, \texttt{LoopVectorizePass} requests loop-structure and target-specific analysis as follows:


\begin{lstlisting}[language=C++,basicstyle=\scriptsize\ttfamily,
  xleftmargin=5pt, 
  emph={LoopVectorizePass,LoopAnalysis,TargetLibraryAnalysis},
  emphstyle=\color{green!70!black}\bfseries]
Preservedanalysis LoopVectorizePass::run(...,
                                AnalysisManager &AM) {
  ...
  auto &LI  = AM.getResult<LoopAnalysis>(F);
  auto &TLI = AM.getResult<TargetLibraryAnalysis>(F);
  ...
}
\end{lstlisting}

Therefore, we obtain the $Deps$ field by locating each pass’s \texttt{run} implementation under \texttt{llvm/lib/ Transforms} and statically parsing all invocations of \texttt{AnalysisManager::getResult}. As shown in \autoref{fig:overview}(b), the dependency set for \texttt{LoopVectorizePass} is \{\texttt{LoopAnalysis}, \texttt{TargetLibraryAnalysis}\}.

\textbf{Retrieve Analysis Results.}
With the constructed compiler knowledge base $\mathcal{K}$, we retrieve the concrete compiler transform passes and their dependent analysis passes corresponding to each transformation action $a_i$ in the initial optimization strategy $S$. Specifically, for each action $a_i$, we treat the natural-language description of $a_i$ as a query and match it against the document corpus $\{Desc(k)\}_{k \in \mathcal{K}}$. Both the query and documents are represented using TF-IDF~\cite{TF-IDF} with 1–3 gram tokenization, and the top-m matching passes are selected based on cosine similarity.
\[
\begin{aligned}
\text{sim}(a_i, k) &= \cos(\text{TF-IDF}(a_i), \text{TF-IDF}(Desc(k))) \\
\{k_i^{(j)}\}_{j=1}^{m} &= \arg\text{top-}m_{k\in\mathcal{K}} \text{sim}(a_i, k)
\end{aligned}
\]
Such design balances efficiency and coverage: top-m retrieval captures cases where multiple passes address similar goals (e.g., \texttt{EarlyCSEPass}, \texttt{DCEPass}, and \texttt{GVNPass} all perform redundancy elimination) while avoiding the computational overhead of indiscriminate analysis execution.

Once the passes $\{k_i^{(j)}\}_{j=1}^{m}$ are identified, we retrieve their required analysis passes  $\textit{Deps}(k_i^{(j)})$ from the knowledge base and invoke the LLVM compiler to generate the corresponding analysis results on the current IR program $P$:
$$R_S = \bigcup_{a_i \in S} \bigcup_{j=1}^{m} \text{LLVM\_Generate}(\textit{Deps}(k_i^{(j)}), P)$$ 
The analysis results are obtained through LLVM’s analysis inspection interfaces (e.g., \texttt{opt -p=print<analysis>}), which generate analysis information in textual form. This information corresponds to the same analysis results that transform passes query during the compiler optimization process to ensure correctness and performance. As a result, they provide reliable information for downstream intent refinement and intent realization.


\textbf{Refinement.} Given the initial optimization strategy $S$ and the retrieved analysis results $R_S$, the refinement stage produces a refined strategy $\hat{S}$:
$$\hat{S} = \mathcal{G}(P, S, R_S)$$
where $\mathcal{G}$ denotes the LLM (GPT-5) that performs intent refinement by reasoning over the program, initial strategy, and compiler analysis information. The prompt template used during refinement is provided in \autoref{app:prompt_template}.


This refinement process validates and enhances the initial strategy in \Cref{sec:intent_formulation} by grounding it in compiler analysis information: it filters infeasible transformations based on program constraints, while leveraging both analysis information and the LLM's pretraining knowledge to refine the transformations in ways that may exceed conventional compiler heuristics. The resulting strategy $\hat{S}$ is constraint-aware and analytically informed, providing a reliable bridge between intent formulation and concrete intent realization.

\subsection{Intent Realization} 
\label{sec:intent_realization}
The intent realization stage finally applies the strategy $\hat{S}$ on the unoptimized IR program $P$ to generate the optimized IR program $\hat{P}$:
$$\hat{P} = \mathcal{G}(P, \hat{S}, R_S)$$
where $\mathcal{G}$ denotes the same LM (GPT-5) used in refinement. This stage synthesizes the optimized program by applying the transformation actions specified in $\hat{S}$ while respecting the constraints encoded in the analysis results $R_S$. The prompt template is provided in \autoref{app:prompt_template}.

Through this three-stage pipeline -- formulation, refinement, and realization -- \system transforms high-level optimization intent into concrete, well-informed IR transformations.  Leveraging GPT-5 for intent refinement and realization allow \system to explore optimization strategies beyond those typically produced by traditional fixed-pass compilers.
\section{Experiments}
This section explores the following research questions:
\vspace{-2ex}
\begin{itemize}[itemsep=2pt, leftmargin=2ex]
\item \textbf{RQ.1} Can \system outperform existing end-to-end LLM-based IR optimization approaches? (\autoref{sec:imp_llmOpt})

\item \textbf{RQ.2} How does \system compare against the traditional compiler in terms of performance? Does the explicit optimization intent lead to performance gains beyond the pass-by-pass paradigm? (\autoref{sec:imp_compiler} \& \autoref{sec:case_study})

\item \textbf{RQ.3} How does each component of \system contribute to its overall results? (\autoref{sec:ablation})
\end{itemize}

\begin{table*}[t] \small
\centering
\caption{Comparison of correctness and performance of six LLM-based IR optimization methods. Correctness is evaluated using Alive2 combined with differential testing, while performance is measured as speedup over the input unoptimized IR.}
\vspace{-2ex}
\label{table:acc_imp}
\vspace{1.ex}
\setstretch{0.95}
\setlength{\tabcolsep}{0.7ex}
\resizebox{0.95\textwidth}{!}{
\begin{tabular}{lcc|cccc}
\hline
\multirow{2}{*}{\textbf{Method}} &
\multicolumn{2}{c|}{\textbf{Correctness}} &
\multicolumn{4}{c}{\textbf{Performance}} \\
\cline{2-7}
& \textbf{Alive2} & \textbf{Alive2+Diff. Test.} & \textbf{Avg. Speedup} &
\textbf{Speedup $> 1.1\times$} &
\textbf{Speedup $> 1.5\times$} &
\textbf{Speedup $> 2.0\times$} \\
\hline
GPT-5           & \cellcolor{green!20} \textbf{73.0\%} (146) & 83.5\% (167) & 1.057$\times$ & 26.0\% (52) & 10.5\% (21) & 6.5\% (13) \\
Claude Haiku 4.5    & 54.0\% (108) & 61.0\% (122) & 0.721$\times$ & 15.5\% (31) & 5.0\% (10) & 2.5\% (5) \\
DeepSeek-V3.2       & 51.0\% (102) & 59.0\% (118) & 0.719$\times$ & 14.5\% (29) & 6.5\% (13) & 3.5\% (7) \\
LLM Compiler FTD 7B     & 57.0\% (114) & 60.5\% (121) & 0.702$\times$ & 13.0\% (26) & 6.0\% (12) & 2.5\% (5) \\
LLM Compiler FTD 13B   & 61.5\% (123) & 64.5\% (129) & 0.757$\times$ & 12.0\% (24) & 4.0\% (8) & 2.5\% (5) \\
\hline
\textbf{\system}   & 71.0\% (142) & \cellcolor{green!20} \textbf{90.5\%} (181) & \cellcolor{green!20} \textbf{2.660$\times$} &
\cellcolor{green!20} \textbf{37.5\%} (75) & \cellcolor{green!20} \textbf{17.5\%} (35) & \cellcolor{green!20} \textbf{10.0\%} (20) \\
\hline
\end{tabular}}
\vspace{-2ex}
\end{table*}

\subsection{Experimental Setup} 
\label{sec:exp_setup}
\subsubsection{Dataset}
We construct all datasets from IR-OptSet~\cite{yang2025iroptset}, an IR-oriented dataset designed to trigger representative optimization transformations encountered in real compiler optimization scenarios, and each IR sample (unoptimized IR plus optimized IR) is capped at 5,000 tokens. We randomly sample 4,000 programs to form the training set \textbf{$\text{INT}_{\text{train}}$} and 500 programs to form the validation set \textbf{$\text{INT}_{\text{val}}$}.
For evaluation, we construct a separate test set \textbf{$\text{INT}_{\text{test}}$} of 200 IR samples through a multi-stage filtering process. First, we retain programs for which Alive2~\cite{alive2} can verify semantic equivalence between the unoptimized IR and LLVM \texttt{-O3}. Second, we exclude programs unsuitable for differential testing~\cite{McKeeman1998DifferentialTF}, such as those with custom external functions, externally defined global variables, or non-deterministic behavior. This ensures that when LLM-based optimizers produce transformations beyond Alive2's verification capability, differential testing can be reliably used for correctness validation.

\subsubsection{Baselines} 
We include two categories of baselines for comparison:

\textbf{End-to-end LLM-based IR optimizers.} We consider both general-purpose and specialized LLM-based optimizers that directly map unoptimized IR to optimized IR. 
Specifically, we evaluate three representative general-purpose LLMs with strong capabilities on code-related tasks: 
\textit{\textbf{GPT-5}}~\cite{Web:gpt5}, 
\textit{\textbf{Claude Haiku 4.5}}~\cite{Web:claude}, and 
\textit{\textbf{DeepSeek-V3.2}}~\cite{deepseekv32}. 
All models are prompted using the template shown in \autoref{app:llm_template}.
We also compare against LLM Compiler~\cite{cummins2025llm}, a state-of-the-art specialized end-to-end IR optimizer, evaluating both the 7B and 13B parameter variants (\textit{\textbf{LLM Compiler FTD 7B}} and \textit{\textbf{LLM Compiler FTD 13B}}). To ensure a fair comparison, both LLM Compiler variants are fine-tuned on $\text{INT}_{\text{train}}$ with $\text{INT}_{\text{val}}$ following \autoref{app:llmcompiler_train}.

\textbf{Compiler baseline.} We include \textbf{\textit{LLVM 19.1.0}}~\cite{Web:LLVM} with the \textbf{\textit{\texttt{-O3}}} optimization level as a compiler-level reference. \texttt{-O3} aggressively optimizes for execution speed~\cite{Web:llvmo3} and represents a highly tuned IR optimization strategy.

\subsubsection{Implementation Details}
For the intent formulation stage of \system, we obtain the optimization-specialized LLM $\mathcal{F}_\theta$ in \system, using the $\text{INT}_{\text{train}}$ with $\text{INT}_{\text{val}}$, following the procedure described in \autoref{sec:intent_formulation}. 
During the analysis results retrieval stage, we select the top-3 matching optimization passes and collect their corresponding LLVM analysis results, which are used to support the intent refinement stage.

\subsubsection{Evaluation Metrics}
\textbf{Correctness Verification.}
We employ a two-stage verification process to validate the \textbf{\textit{Correctness}} of optimized IR programs. We first apply Alive2 to perform strict formal semantic equivalence checking between the optimized and unoptimized IR. Since Alive2 may time out on verifying programs with complex control flow or memory behaviors~\cite{alive2}, we additionally apply differential testing when formal verification is infeasible. Specifically, we use LibFuzzer~\cite{Web:libFuzzer} to generate diverse test inputs and verify that optimized and unoptimized programs produce identical outputs across different execution paths. Details are provided in \autoref{app:verification}.

\textbf{Performance Evaluation.}
We measure performance using the same inputs generated by LibFuzzer during correctness verification, reflecting execution behaviors across diverse control flow paths. Each program is executed for 10,000 iterations to obtain stable timing measurements and mitigate measurement noise. 
We report \textbf{\textit{Performance}} as speedup ratio: 
$\text{Speedup} = \text{RunTime(unopt)}/\text{RunTime(opt)}$.
When an LLM-based optimizer generates incorrect IR, we assign a speedup of 0. More details are in ~\autoref{app:performance}. 

\begin{figure*}
    \centering
    \includegraphics[width=0.9\linewidth]{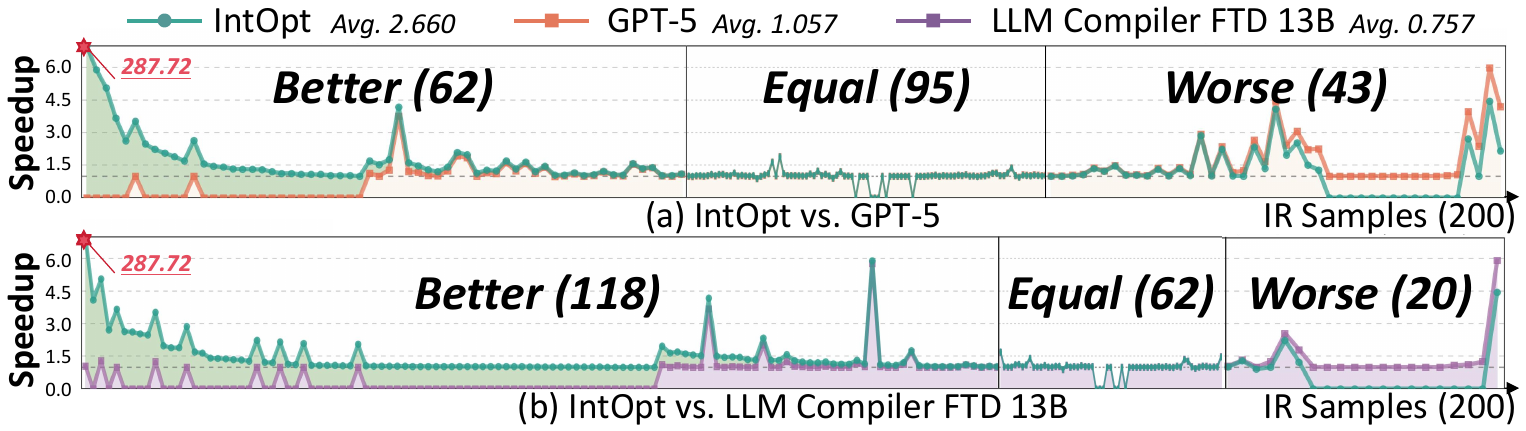}
    \vspace{-2ex}
    \caption{Per-program speedup comparison between \system and representative end-to-end LLM-based IR optimizers. Speedup differences within $\pm$2\% are considered equal to account for runtime variability.}
    \label{fig:perf_imp}
    \vspace{-3ex}
\end{figure*}

\begin{figure*}
    \centering
    \includegraphics[width=0.85\linewidth]{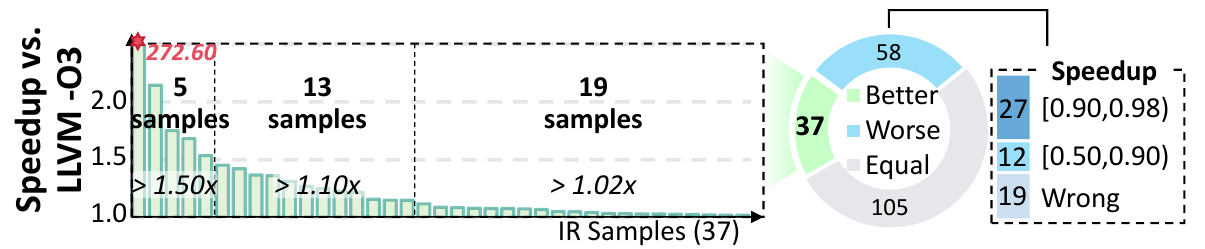}
    \vspace{-1ex}
    \caption{
    Performance comparison between \system and LLVM \texttt{-O3}. 
    Speedup is defined as $\text{RunTime}(\text{LLVM \texttt{-O3}}) / \text{RunTime}(\text{opt})$, with $[0.98,1.02]$ treated as equal. 
    \system matches LLVM \texttt{-O3} on 105 programs and outperforms it on 37 (max $272.60\times$); among 58 underperforming cases, most are slight slowdowns (27 in $[0.90,0.98]$), and 19 are incorrect.
    }
    \label{fig:perf_imp_o3}
    \vspace{-2ex}
\end{figure*}

\subsection{Improvement over LLM-based IR Optimizers}
\label{sec:imp_llmOpt}

We first compare \system with end-to-end LLM-based IR optimizers on the 200-sample benchmark $\text{INT}_{\text{test}}$. As shown in ~\autoref{table:acc_imp}, \system achieves the strongest overall results across all baselines in both correctness and performance.

\textbf{Correctness.} Under Alive2-only verification, GPT-5 achieves the highest correctness rate among baselines (73.0\%), while \system attains a comparable rate of 71.0\%. However, this metric is inherently limited by Alive2’s verification capability: even complex optimizations produced by LLVM \texttt{-O3} cannot always be verified~\cite{alive2}. When differential testing is incorporated, \system achieves a verified correctness rate of 90.5\%, surpassing all other baselines by 7.0\%-31.5\%. These results highlight \system’s advantage in reducing semantic errors over end-to-end LLM-based IR optimization methods.

\textbf{Performance.}
\system achieves an average speedup of 2.660$\times$ over the unoptimized IR, significantly outperforming all end-to-end LLM-based baselines.The strongest baseline, GPT-5, attains an average speedup of only 1.057$\times$, resulting in a 1.603$\times$ performance gap.
Beyond average performance, \system also consistently dominates across all speedup thresholds, indicating that its performance gains are broadly distributed across the benchmark.

To better understand these improvements, we further compare \system against the two strongest baselines, GPT-5 and LLM Compiler FTD 13B, on a per-program basis, as shown in \autoref{fig:perf_imp}.
Against GPT-5, \system achieves higher speedups on 62 programs, while the advantage is more pronounced against LLM Compiler FTD 13B, where \system outperforms it on 118 programs.
Notably, when compared to the unoptimized IR, \system attains a maximum speedup of up to $287.72\times$ on a single program.
These results suggest that \system is more effective at discovering and realizing aggressive yet profitable optimization opportunities.

\colorbox{gray!10}{
\parbox{\linewidth}{
\underline{\textit{\textbf{Answer to RQ.1:}}} By decoupling optimization intent formulation from concrete transformation generation, \system reduces semantic errors and is more effective at discovering profitable optimization opportunities than existing end-to-end LLM-based IR optimization approaches.
}}

\vspace{-2ex}
\subsection{Improvement over Traditional Compiler}
\label{sec:imp_compiler}

We further compare \system's performance against LLVM \texttt{-O3}, a highly optimized and widely adopted compiler optimization level.
\autoref{fig:perf_imp_o3} shows the comparison results.

\system achieves performance comparable to LLVM \texttt{-O3} on 105 programs and outperforms it on \textbf{37} programs.
Among these improved cases, the per-program speedup distribution reveals a spectrum of performance gains: 5 programs exhibit substantial improvements with speedups greater than $1.5\times$, while 13 programs achieve clear improvements exceeding $1.1\times$.
Notably, the program that achieves the highest speedup in \autoref{fig:perf_imp} also attains the highest performance improvement over LLVM \texttt{-O3}, with a speedup of up to $272.60\times$.
We analyze this case in detail in \autoref{sec:case_study}, illustrating how \system can uncover optimization opportunities that go beyond the capabilities of the traditional compiler.

\begin{figure*}
    \centering
    \includegraphics[width=0.95\linewidth]{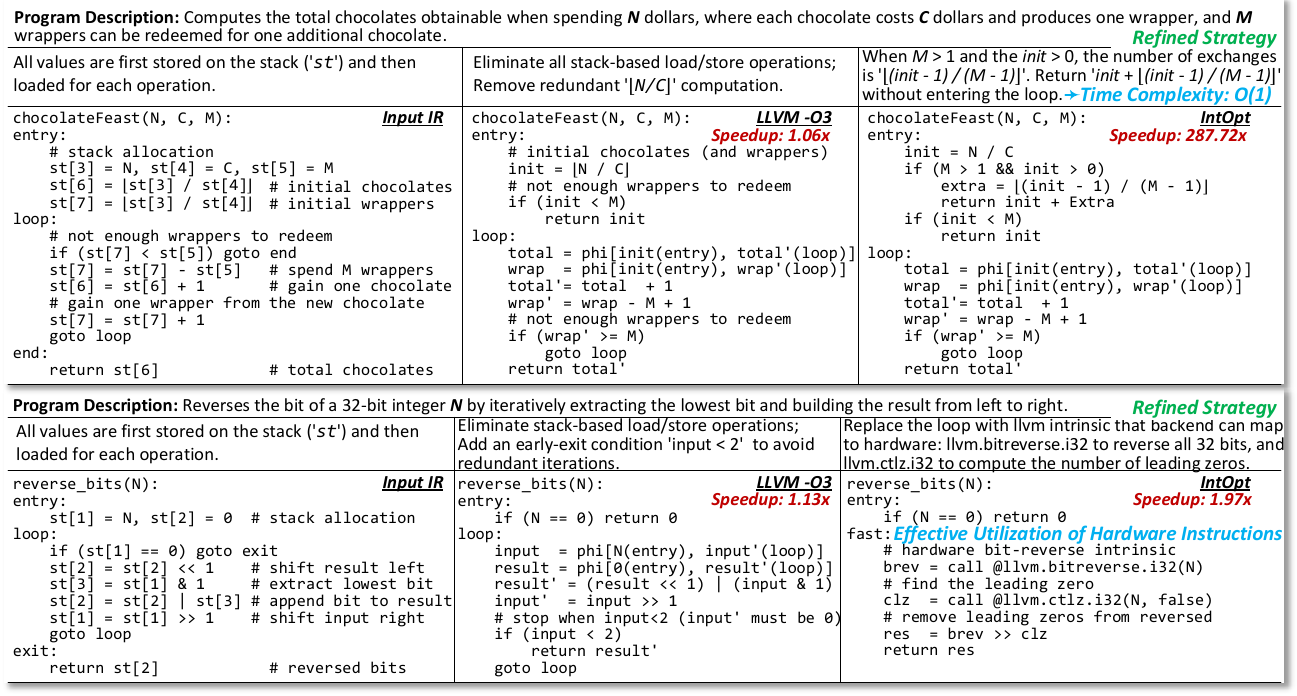}
    \vspace{-1ex}
    \caption{
    Case study illustrating how \system explores optimization opportunities beyond the traditional pass-by-pass compiler paradigm.
    }
    \label{fig:case_study}
\end{figure*}

\begin{table*}[t] \small
\centering
\caption{Ablation study of \system’s design. Impact of intent refinement and compiler analysis on correctness and performance, where performance is measured as speedup over the input unoptimized IR.}
\label{table:ablation}
\setstretch{0.95}
\setlength{\tabcolsep}{0.7ex}
\resizebox{0.95\textwidth}{!}{
\begin{tabular}{lcc|cccc}
\hline
\multirow{2}{*}{\textbf{Method}} &
\multicolumn{2}{c|}{\textbf{Correctness}} &
\multicolumn{4}{c}{\textbf{Performance}} \\
\cline{2-7}
& \textbf{Alive2} & \textbf{Alive2+Diff. Test.} & \textbf{Avg. Speedup} &
\textbf{Speedup $> 1.1\times$} &
\textbf{Speedup $> 1.5\times$} &
\textbf{Speedup $> 2.0\times$} \\
\hline
\textbf{\system}   & 71.0\% (142) & \textbf{90.5\%} (181) & \textbf{2.660$\times$} &
\textbf{37.5\%} (75) & \textbf{17.5\%} (35) & \textbf{10.0\%} (20) \\
\quad w/o Intent Refinement & \textbf{71.5\%} (143) & 87.0\% (174) & 1.122$\times$ &
30.0\% (60) & 15.0\% (30) & 7.5\% (15) \\
\quad w/o Analysis & 65.0\% (130) & 77.0\% (154) & 2.415$\times$ &
29.0\% (58) & 13.0\% (26) & 6.5\% (13) \\

\hline
\end{tabular}}
\vspace{-2ex}
\end{table*}


\subsection{Case Analysis}
\label{sec:case_study} 

To provide deeper insights into the intent-driven optimization mechanism, we present two representative cases that illustrate how \system explores optimization opportunities beyond the traditional pass-by-pass compiler paradigm.
Due to space constraints, we present pseudo-code representations of the IR; the full code listings are provided in \autoref{app:case_study}.
As shown in \autoref{fig:case_study}, traditional compilers primarily optimize programs by matching pre-defined patterns in passes designed by compiler developers and applying transformations (e.g., eliminating stack-based load/store operations).

In contrast, \system performs intent-guided optimization that enables coordinated,
semantic-aware transformations.
In \texttt{chocolateFeast}, \system identifies an algorithmic shortcut during intent refinement that reduces the time complexity of an iterative computation to \(O(1)\) and guides the subsequent optimization, achieving a $287.72\times$ speedup.
In \texttt{reverse\_bits}, \system achieves a $1.97\times$ speedup by coordinating multiple transformations.
After recognizing the high-level semantics of the loop as bit reversal, \system replaces the original computation with the \texttt{llvm.bitreverse} intrinsic, and further applies an auxiliary transformation using \texttt{llvm.ctlz} with shift-based operations to model the original loop's exit condition.
These optimizations are difficult for existing compilers to realize, as they require semantic reasoning beyond what local pattern matching in a fixed pass-by-pass pipeline can capture.


\colorbox{gray!10}{
\parbox{\linewidth}{
\underline{\textit{\textbf{Answer to RQ.2:}}} Explicit optimization intent enables \system to realize optimizations that go beyond the capabilities of traditional pass-by-pass compilers.
}}

\subsection{Effectiveness of \system’s Design}
\label{sec:ablation}

The intent formulation stage is a necessary first stage in this pipeline, while intent realization serves as the final stage that generates optimized IR; neither can be bypassed.
Accordingly, our ablation study focuses on the contribution of the intent refinement stage and the impact of compiler analysis on \system.
As shown in \autoref{table:ablation}, removing intent refinement slightly improves Alive2-only correctness but leads to a noticeable degradation in both verified correctness and performance, reducing average speedup from $2.660\times$ to $1.122\times$. Removing compiler analysis from both refinement and realization (\emph{w/o Analysis}) further reduces verified correctness to 77.0\%, as refinement and realization decisions are made without sufficient semantic constraints. Despite this, performance remains relatively high ($2.415\times$ on average). This behavior can be attributed to the fact that, starting from compiler-derived optimization intents produced during intent formulation, the LLM can occasionally identify highly aggressive transformations, including the case shown in \autoref{fig:perf_imp} that achieves a speedup of $287.72\times$.



\colorbox{gray!10}{
\parbox{\linewidth}{
\underline{\textit{\textbf{Answer to RQ.3:}}} Both intent refinement and compiler analysis are both critical to \system, enabling effective optimization discovery and reliable realization.
}}

%





\section{Conclusion}

This paper introduces \system, the first intent-driven IR optimization framework. By decoupling high-level optimization intent from low-level analysis and transformation, \system addresses a critical yet underexplored dimension of IR optimization. Our experimental results show that \system delivers substantial improvements over both end-to-end LLM-based optimizers and traditional compiler optimizations, demonstrating that making optimization intent explicit is crucial for achieving globally coordinated, semantically sound, and high-performance IR optimization.




\section*{Impact Statement}

This work explores the use of LLMs to assist IR optimization in compilers, a domain where correctness, reliability, and performance are critical. By explicitly modeling optimization intent and separating it from low-level transformations, the proposed framework enhances the interpretability and inspectability of LLM-assisted compiler optimization, enabling developers to better understand and verify the optimization strategies produced by these systems. We believe that intent-driven optimization provides a promising foundation for future research at the intersection of compilers and machine learning, fostering more trustworthy and explainable AI-assisted program optimization.




\bibliography{custom}
\bibliographystyle{icml2026}

\newpage
\appendix
\onecolumn

\section{Example of Compiler Analysis Information}

In this section, we provide a detailed example of how compiler analysis information guides the optimization process for an input IR program, as shown in \autoref{fig:analysis_exmp_input}. It contains two types of branches: (1) unconditional branch (e.g., \texttt{br label \%B1}), which is a type of branch directly jumps to another block without any conditions; (2) conditional branch (\texttt{br i1 \%16, label \%B2, label \%B3}), which is a type of branch performs a jump based on a condition.

\begin{figure}[H]
    \begin{lstlisting}[
    frame=single,
    rulecolor=\color{black},
    backgroundcolor=\color{gray!10},
    basicstyle=\ttfamily\small,
    breaklines=true,
    % breakindent=0pt,
    breakautoindent=false,
    showstringspaces=false
    ]
define dso_local i32 @chocolateFeast(i32 noundef %0, i32 noundef %1, i32 noundef %2){
B0:
%3 = alloca i32, align 4
%4 = alloca i32, align 4
%5 = alloca i32, align 4
%6 = alloca i32, align 4
%7 = alloca i32, align 4
store i32 %0, ptr %3, align 4, !tbaa !5
store i32 %1, ptr %4, align 4, !tbaa !5
store i32 %2, ptr %5, align 4, !tbaa !5
call void @llvm.lifetime.start.p0(i64 4, ptr %6) #2
%8 = load i32, ptr %3, align 4, !tbaa !5
%9 = load i32, ptr %4, align 4, !tbaa !5
%10 = sdiv i32 %8, %9
store i32 %10, ptr %6, align 4, !tbaa !5
call void @llvm.lifetime.start.p0(i64 4, ptr %7) #2
%11 = load i32, ptr %3, align 4, !tbaa !5
%12 = load i32, ptr %4, align 4, !tbaa !5
%13 = sdiv i32 %11, %12
store i32 %13, ptr %7, align 4, !tbaa !5
br label %B1
B1:
%14 = load i32, ptr %5, align 4, !tbaa !5
%15 = load i32, ptr %7, align 4, !tbaa !5
%16 = icmp sle i32 %14, %15
br i1 %16, label %B2, label %B3
B2:
%17 = load i32, ptr %7, align 4, !tbaa !5
%18 = load i32, ptr %5, align 4, !tbaa !5
%19 = sub nsw i32 %17, %18
store i32 %19, ptr %7, align 4, !tbaa !5
%20 = load i32, ptr %6, align 4, !tbaa !5
%21 = add nsw i32 %20, 1
store i32 %21, ptr %6, align 4, !tbaa !5
%22 = load i32, ptr %7, align 4, !tbaa !5
%23 = add nsw i32 %22, 1
store i32 %23, ptr %7, align 4, !tbaa !5
br label %B1, !llvm.loop !9
B3:
%24 = load i32, ptr %6, align 4, !tbaa !5
call void @llvm.lifetime.end.p0(i64 4, ptr %7) #2
call void @llvm.lifetime.end.p0(i64 4, ptr %6) #2
ret i32 %24
}
    \end{lstlisting}
    \caption{Example of the input IR}
    \label{fig:analysis_exmp_input}
\end{figure}

Below, we break down the key elements of the analysis and explain what each part means.

\subsection{Dominator Tree Analysis}
The Dominator Tree is a data structure used to identify the dominance relationships between basic blocks in a control flow graph (CFG) of the function. A block \texttt{B} dominates another block \texttt{C} if every path from the entry block to \texttt{C} must go through \texttt{B}. The Dominator Tree helps identify potential places for optimization by highlighting which blocks are crucial for execution and which may be redundant or amenable to optimization. For the function \texttt{chocolateFeast}, the dominator tree is shown as follows:

\begin{lstlisting}[
    frame=single,
    rulecolor=\color{black},
    backgroundcolor=\color{gray!10},
    basicstyle=\ttfamily\small,
    breaklines=true,
    % breakindent=0pt,
    breakautoindent=false,
    showstringspaces=false
    ]
DominatorTree for function: chocolateFeast
=============================--------------------------------
Inorder Dominator Tree: DFSNumbers invalid: 0 slow queries.
  [1] %B0
    [2] %B1
      [3] %B2
      [3] %B3
Roots: %B0 
\end{lstlisting}

\begin{itemize}
    \item \texttt{\%B0} is at level 1, which indicates it is the entry block (starting point of the function). As the root of the dominator tree, \texttt{\%B0} dominates all other blocks. It is the first block that will be executed when the function is called, and every other block must pass through it.

    \item \texttt{\%B1} is at level 2, meaning it is one step deeper in the control flow compared to \texttt{\%B0}. This block is dominated by \texttt{\%B0}, and it is the first block executed after \texttt{\%B0}.
    
    \item Both \texttt{\%B2} and \texttt{\%B3} are at level 3, meaning they are two steps deeper in the control flow than \texttt{\%B0}. These blocks are dominated by \texttt{\%B1}, meaning they are only executed after \texttt{\%B1}.
\end{itemize}

\subsection{Loop Analysis}

Loops are a key target for optimization, as they are often the source of performance bottlenecks. Loop analysis identifies the loops within the function and provides details on their structure, such as their depth and the basic blocks involved. In a loop structure, the loop header is the first block in the loop, where all edges from outside the loop converge. The loop latch is a node inside the loop with an edge pointing back to the loop header. The exit block is the block where control exits the loop, marking the termination point of the loop.
For the function \texttt{chocolateFeast}, the loop analysis is shown as follows:

\begin{lstlisting}[
    frame=single,
    rulecolor=\color{black},
    backgroundcolor=\color{gray!10},
    basicstyle=\ttfamily\small,
    breaklines=true,
    % breakindent=0pt,
    breakautoindent=false,
    showstringspaces=false
    ]
Loop info for function 'chocolateFeast':
Loop at depth 1 containing: %B1<header><exiting>,%B2<latch>
\end{lstlisting}

\begin{itemize}
    \item The loop is at depth 1, meaning it's a top-level loop in the function.
    \item The loop involves \texttt{\%B1} as the loop header and the exit block, \texttt{\%B2} as the loop latch.
\end{itemize}
\section{Prompt Template for \system}
\label{app:prompt_template}

\subsection{Intent Formulation Prompt Template}

The intent formulation stage in \system generates an initial optimization strategy $S$, which serves as a structured instantiation of the optimization intent for a given input IR program $P$. To train the fine-tuned LLM Compiler FTD 13B, we use the following template to guide it in producing an ordered sequence of natural language transformation actions:

\begin{lstlisting}[
frame=single,
rulecolor=\color{black},
backgroundcolor=\color{gray!10},
basicstyle=\ttfamily\small,
breaklines=true,
breakindent=0pt,
breakautoindent=false,
showstringspaces=false
]
[INST]Given the following LLVM IR, propose key optimization transformation steps to outperform LLVM -O3.
Write your answer inside a single <code>...</code> block.
Inside <code>, write ONLY <step></step> blocks.
Each step MUST follow this format:
<step>
**Transformation**: [Brief name of the optimization]
**Change**: [A short description of the change applied to the code]
</step>
Do NOT output optimized IR.

<ir>(Unopt Input LLVM IR)</ir>
[\INST]
\end{lstlisting}

\subsection{Intent Refinement Prompt Template}

The intent refinement stage takes the initial optimization strategy $S$, which specifies a sequence of transformations for a given input IR program $P$, and refines it by grounding the strategy in concrete compiler analysis results $R_S$. The template used in intent refinement stage is shown as follows:

\begin{lstlisting}[
frame=single,
rulecolor=\color{black},
backgroundcolor=\color{gray!10},
basicstyle=\ttfamily\small,
breaklines=true,
breakindent=0pt,
breakautoindent=false,
showstringspaces=false
]
Please optimize the following code to outperform LLVM -O3.
<code>(Unopt LLVM IR)</code>

You may refer to the following advice, but feel free to adapt, extend, or deviate from it as you see fit.
<advice>(Initial Strategy)</advice>

The corresponding analysis info is below.
<analysis>(Compiler Analysis)</analysis>

You need to keep boundary checks. Please output the final optimization advice wrapped in <advice>...</advice>.
\end{lstlisting}

\subsection{Intent Realization Prompt Template}

The intent realization stage takes the refined optimization strategy $\hat{S}$ and applies it to the unoptimized IR program $P$ to generate the optimized IR program $\hat{P}$. The template used in intent realization stage is shown as follows:

\begin{lstlisting}[
frame=single,
rulecolor=\color{black},
backgroundcolor=\color{gray!10},
basicstyle=\ttfamily\small,
breaklines=true,
breakindent=0pt,
breakautoindent=false,
showstringspaces=false
]
Please optimize the following code to outperform LLVM -O3.
<code>(Unopt LLVM IR)</code>

You can refer to the following advice.
<advice>(Refined Strategy)</advice>

The corresponding analysis info is below.
<analysis>(Compiler Analysis)</analysis>

You need to keep boundary checks. Please output the full optimized LLVM IR wrapped in <code>...</code>.
\end{lstlisting}

\section{Prompt Template for LLM-based Optimizers}
\label{app:llm_template}

To evaluate the performance of general-purpose LLM-based optimizers (GPT-5, Claude Haiku 4.5, and DeepSeek-V3.2), we use the following prompt template to guide the optimization process:

\begin{lstlisting}[
frame=single,
rulecolor=\color{black},
backgroundcolor=\color{gray!10},
basicstyle=\ttfamily\small,
breaklines=true,
breakindent=0pt,
breakautoindent=false,
showstringspaces=false
]
Please optimize the following code to outperform LLVM -O3.
<code>(Unopt LLVM IR)</code>

You need to keep boundary checks. Please output the full optimized LLVM IR wrapped in <code>...</code>.
\end{lstlisting}
\section{Token Statistic}

\autoref{tab:token_stats} shows the data statistics about the number and token of $\text{INT}_{\text{train}}$, $\text{INT}_{\text{val}}$, and $\text{INT}_{\text{test}}$ used in \autoref{sec:exp_setup}. Specifically, we generated the dataset for optimization intent using $\text{INT}{\text{train}}$ and $\text{INT}_{\text{val}}$ following the procedure described in \autoref{sec:intent_formulation}.

\begin{table}[H] \small
\centering
\caption{Data statistics about the number and token of $\text{INT}_{\text{train}}$, $\text{INT}_{\text{val}}$, and $\text{INT}_{\text{test}}$.}
\label{tab:token_stats}
\setstretch{0.95}
\begin{tabular}{lllllll}
\toprule
                            &                     & \textbf{Total Token} & \textbf{Average Token} & \textbf{Median Token} & \textbf{Min Token} & \textbf{Max Token} \\
\midrule
\multirow{3}{*}{\textbf{$\text{INT}_{\text{train}}$}} & Unoptimized IR      &  8,219,999           &   2,055.0            &   2,047.0           &   579        &     4,304      \\
                            & Optimized IR        &   5,531,194          &   1,382.8            &  1,370.5            &   359        &     3,373      \\
                            & Optimization Intent &  3,753,392           &   938.6            &   933.0           &     331      &     1,719      \\
\midrule
\multirow{3}{*}{\textbf{$\text{INT}_{\text{val}}$}} & Unoptimized IR      &  1,017,905           &   2,035.8            &  2016.0            &    685       &    4,035       \\
                            & Optimized IR        &   691,180          &    1,382.4           &   1,359.5           &    425       &      3,220     \\
                            & Optimization Intent &    473,842         &    947.7           &     941.5         &    321       &     1,692      \\
\midrule
\textbf{$\text{INT}_{\text{test}}$}       & Unoptimized IR      &   362,455          &   1,812.3            &   1,572.5           &     679      &      4,188     \\
\bottomrule
\end{tabular}
\end{table}



\section{Experimental Details}
\label{app:llmcompiler_train}

In \autoref{tab:lora_settings}, we provide all hyperparameter settings used in our experiments, including those for fine-tuning LLM Compiler FTD 7B and 13B for IR optimization, as well as fine-tuning LLM Compiler FTD 13B for intent formulation. In this work, we set the context window to 5K, meaning the total input and output must not exceed 5K due to resource limitations. 

\begin{table}[H]
  \small
  \centering
  \caption{Hyperparameter settings.}
  \label{tab:lora_settings}
  \vspace{1ex}
  \setstretch{1}
    \begin{tabular}{lccccccc}
    \toprule
    \textbf{Model} 
      & \thead{LoRA\\rank ($r$)} 
      & \thead{LoRA\\$\alpha$} 
      & \thead{LoRA\\dropout} 
      & \thead{Batch\\size} 
      & \thead{Learning\\rate} 
      & \textbf{Target modules} \\
    \midrule
    LLM Compiler FTD 13B/7B 
      & 48 & 16 & 0.05 & 2 
      & $1\times10^{-4}$ 
      & \makecell[l]{\texttt{\{q\_proj, k\_proj,} \\ \texttt{v\_proj, o\_proj\}}} \\

    \bottomrule
  \end{tabular}
\end{table}

\section{Correctness Validation via Differential Testing}
\label{app:verification}

Differential testing, also known as differential fuzzing, is a software testing technique that detects bugs by providing the same input to a series of similar applications (or to different implementations of the same application), and observing differences in their execution. Differential testing complements traditional software testing because it is well-suited to find semantic or logic bugs that do not exhibit explicit erroneous behaviors like crashes or assertion failures. In this work, we employ libFuzzer to generate a variety of inputs that cover diverse control flow paths, ensuring comprehensive testing of function behavior.

\subsection{Harness Generation for Differential Testing}
\label{app:fuzz-exp}
We leverage GPT-5 to automatically generate libFuzzer harnesses for each function pair. Specifically, for each LLVM IR file containing both a base function f and its optimized variant f\_opt, we prompt the LLM to synthesize a C++ fuzzing harness (fuzz.cc) using the following prompt template:

\begin{lstlisting}[
frame=single,
rulecolor=\color{black},
backgroundcolor=\color{gray!10},
basicstyle=\ttfamily\small,
breaklines=true,
breakindent=0pt,
breakautoindent=false,
showstringspaces=false
]
You are generating a C++ libFuzzer harness file named fuzz.cc.

Input:
- A single LLVM IR (.ll) file contains multiple function definitions.
- Functions that should be differentially fuzzed appear as pairs:
  - base: <name>
  - opt : <name>_opt
  Both have identical signatures.
- The .ll file will be compiled and linked together with fuzz.cc into one binary.

Task:
Generate fuzz.cc that performs differential fuzzing between each (base,opt) pair.

Hard requirements:
1) Output ONLY valid C++ code for fuzz.cc. No markdown fences. No explanations.
2) Include necessary #includes.

LLVM IR content: {ll_text}
\end{lstlisting}

To illustrate this, consider the following example of a C++ fuzzing harness generated for a function pair \texttt{numberOfOperations} and \texttt{numberOfOperations\_opt}:

\begin{lstlisting}[
frame=single,
rulecolor=\color{black},
backgroundcolor=\color{gray!10},
basicstyle=\ttfamily\small,
breaklines=true,
breakindent=0pt,
breakautoindent=false,
showstringspaces=false
]
#include <cstdint>
#include <cstddef>
#include <cstring>

int numberOfOperations(int);
int numberOfOperations_opt(int);

extern "C" int LLVMFuzzerTestOneInput(const uint8_t* data, size_t size) {
    if (size < 4) return 0;

    int32_t in32;
    std::memcpy(&in32, data, 4);
    int arg = static_cast<int>(in32);

    int r_base = numberOfOperations(arg);
    int r_opt  = numberOfOperations_opt(arg);

    if (r_base != r_opt) {
        __builtin_trap();
    }

    return 0;
}
\end{lstlisting}

The function \texttt{LLVMFuzzerTestOneInput} is the entry point for libFuzzer, receiving input data generated by libFuzzer. The data is first parsed as a 32-bit integer, which is then passed as an argument to both \texttt{numberOfOperations} and \texttt{numberOfOperations\_opt}. If the results of the base and optimized functions do not match, the program triggers a trap using \texttt{\_\_builtin\_trap()}, halting execution and indicating a mismatch.

\subsection{Differential Testing Binary Generation and Execution}

Each generated harness is compiled and linked with its corresponding LLVM IR to produce an instrumented fuzzing binary. The compilation pipeline consists of:

\begin{itemize}
    \item \textbf{IR Compilation.} Since \texttt{llc} operates on the backend, this step applies target-specific code generation and backend optimizations while preserving the IR-level transformations produced by the different optimizer.
    \item \textbf{Sanitized Linking.} The generated assembly is linked with the fuzzing harness (\texttt{fuzz.cc}) using \texttt{clang++}, with comprehensive sanitizer instrumentation enabled via \texttt{-fsanitize=fuzzer,address,undefined}. These sanitizers respectively support coverage-guided input generation, detection of memory safety violations (e.g., buffer overflows and use-after-free), and identification of undefined behavior inconsistencies between the base and optimized implementations.
\end{itemize}

Each instrumented binary is executed with 200,000 fuzzing iterations (-runs=200000). During execution, LibFuzzer continuously generates diverse test inputs guided by code coverage feedback, invoking both the base and optimized functions with identical inputs and comparing their behaviors. Any discrepancy in outputs, crashes, or sanitizer violations between the base and optimized functions indicates a semantic-altering transformation, thereby invalidating the optimization candidate. Only function pairs that successfully complete all 200,000 runs without detecting behavioral differences are validated as semantically equivalent optimizations.
\section{Performance Validation}
\label{app:performance}

We measure performance using the same inputs generated by libFuzzer during correctness verification, thereby reflecting execution behavior across diverse control-flow paths. Based on this design, we transform validated differential fuzzing harnesses into standalone microbenchmarks for performance measurement.

Specifically, we automatically derive microbenchmark harnesses from the validated fuzzing harnesses through a sequence of source-level transformations. Starting from a differential fuzzing harness that invokes a base function \texttt{f} and its optimized variant \texttt{f\_opt}, we first identify the corresponding function symbols (e.g., \texttt{numberOfOperations} and \texttt{numberOfOperations\_opt} in \autoref{app:fuzz-exp}). We then rename the fuzzing entry point \texttt{LLVMFuzzerTestOneInput} to a standard helper function \texttt{decode\_input}, which continues to parse raw input bytes into concrete arguments and performs a single invocation of both implementations.

Next, we inject fine-grained timing instrumentation into \texttt{decode\_input}. As illustrated in the example, we use \texttt{clock\_gettime(CLOCK\_MONOTONIC\_RAW)} to record timestamps immediately before and after each call to the base and optimized functions. Execution time and invocation counts are accumulated in global counters (e.g., \texttt{g\_t\_baseline\_ns}, \texttt{g\_t\_opt\_ns}). To prevent dead-code elimination, the addresses of return values are folded into a global volatile sink variable (\texttt{g\_sink}), ensuring that both function calls are preserved by the compiler.

Finally, we synthesize a standalone \texttt{main} function to drive benchmarking. This driver reads concrete inputs from the fuzzing corpus, performs a fixed number of warmup iterations (e.g., 1{,}000 executions excluded from measurement), and then executes a large number of timed iterations (default: 1{,}000{,}000). After execution, it reports the per-call average runtime of both the base and optimized implementations, along with the resulting speedup. 

All microbenchmarks are executed on an Intel(R) Xeon(R) Gold 6434 processor, and on average, each microbenchmark is evaluated using 16.6 distinct inputs from the fuzzing corpus.

\begin{lstlisting}[
frame=single,
rulecolor=\color{black},
backgroundcolor=\color{gray!10},
basicstyle=\ttfamily\small,
breaklines=true,
breakindent=0pt,
breakautoindent=false,
showstringspaces=false
]
#include <cstdint>
#include <cstddef>
#include <cstring>

#include <time.h>
#include <fstream>
#include <iostream>
#include <vector>
#include <cstdint>

static inline uint64_t now_ns() {
    struct timespec ts;
    clock_gettime(CLOCK_MONOTONIC_RAW, &ts);
    return (uint64_t)ts.tv_sec * 1000000000ull + (uint64_t)ts.tv_nsec;
}

static uint64_t g_t_baseline_ns = 0;
static uint64_t g_t_opt_ns      = 0;
static uint64_t g_n_baseline    = 0;
static uint64_t g_n_opt         = 0;

static volatile uintptr_t g_sink = 0;

int numberOfOperations(int) asm("_Z18numberOfOperationsi");
int numberOfOperations_opt(int) asm("_Z18numberOfOperationsi_opt");
static int decode_input(const uint8_t* data, size_t size) {
    if (size < 4) return 0;

    int32_t in32;
    std::memcpy(&in32, data, 4);
    int arg = static_cast<int>(in32);

    uint64_t __t0 = now_ns();

    int r_base = numberOfOperations(arg);

    uint64_t __t1 = now_ns();

    g_t_baseline_ns += (__t1 - __t0);

    g_n_baseline++;

    g_sink ^= (uintptr_t)(const void*)&r_base;
    uint64_t __t2 = now_ns();
    int r_opt  = numberOfOperations_opt(arg);
    uint64_t __t3 = now_ns();
    g_t_opt_ns += (__t3 - __t2);
    g_n_opt++;
    g_sink ^= (uintptr_t)(const void*)&r_opt;
    if (r_base != r_opt) {
        __builtin_trap();
    }

    return 0;
}

static std::vector<uint8_t> read_file(const char* path) {
    std::ifstream ifs(path, std::ios::binary);
    if (!ifs) {
        std::cerr << "Failed to open: " << path << "\n";
        exit(1);
    }
    return std::vector<uint8_t>(
        (std::istreambuf_iterator<char>(ifs)),
        std::istreambuf_iterator<char>()
    );
}

int main(int argc, char** argv) {
    if (argc < 2) {
        std::cerr << "Usage: " << argv[0] << " <corpus_file> [iters]\n";
        return 1;
    }
    uint64_t iters = (argc >= 3) ? strtoull(argv[2], nullptr, 10) : 1000000ull;

    auto data = read_file(argv[1]);
    if (data.empty()) return 0;

    // donnot measure the warmup time
    uint64_t sb=g_t_baseline_ns, so=g_t_opt_ns, nb=g_n_baseline, no=g_n_opt;
    for (int i = 0; i < 1000; i++) (void)decode_input(data.data(), data.size());
    g_t_baseline_ns=sb; g_t_opt_ns=so; g_n_baseline=nb; g_n_opt=no;

    for (uint64_t i = 0; i < iters; i++) {
        (void)decode_input(data.data(), data.size());
    }

    double avg_b = g_n_baseline ? (double)g_t_baseline_ns / (double)g_n_baseline : 0.0;
    double avg_o = g_n_opt      ? (double)g_t_opt_ns      / (double)g_n_opt      : 0.0;

    std::cout << "iters=" << iters << "\n";
    std::cout << "baseline calls=" << g_n_baseline << " avg(ns/call)=" << avg_b << "\n";
    std::cout << "opt      calls=" << g_n_opt      << " avg(ns/call)=" << avg_o << "\n";
    if (avg_o > 0) std::cout << "speedup=" << (avg_b / avg_o) << "x\n";
    std::cout << "(ignore) sink=" << (unsigned long long)g_sink << "\n";
    return 0;
}

\end{lstlisting}
\section{Case Study}
\label{app:case_study}

\subsection{Algorithm-level Optimization}

\textbf{Program Description.} The \texttt{chocolateFeast} function implements a chocolate purchasing and wrapper exchange simulation. Given an initial budget \texttt{n}, chocolate price \texttt{c}, and exchange rate \texttt{m} (number of wrappers needed to exchange for one chocolate), it calculates the total number of chocolates that can be consumed. The algorithm proceeds in two phases: (1) initial purchase using the budget, and (2) iterative wrapper exchanges until insufficient wrappers remain.

\textbf{Baseline Implementation (Unoptimized).} The unoptimized IR implements this logic using explicit memory allocations, repeated loads and stores, and a loop that simulates wrapper exchanges one iteration at a time.

\begin{lstlisting}[
frame=single,
rulecolor=\color{black},
backgroundcolor=\color{gray!10},
basicstyle=\ttfamily\small,
breaklines=true,
breakindent=0pt,
breakautoindent=false,
showstringspaces=false
]
define dso_local i32 @chocolateFeast(i32 noundef %0, i32 noundef %1, i32 noundef %2) #0 {
B0:
%3 = alloca i32, align 4
%4 = alloca i32, align 4
%5 = alloca i32, align 4
%6 = alloca i32, align 4
%7 = alloca i32, align 4
store i32 %0, ptr %3, align 4
store i32 %1, ptr %4, align 4
store i32 %2, ptr %5, align 4
call void @llvm.lifetime.start.p0(i64 4, ptr %6) #2
%8 = load i32, ptr %3, align 4
%9 = load i32, ptr %4, align 4
%10 = sdiv i32 %8, %9
store i32 %10, ptr %6, align 4
call void @llvm.lifetime.start.p0(i64 4, ptr %7) #2
%11 = load i32, ptr %3, align 4
%12 = load i32, ptr %4, align 4
%13 = sdiv i32 %11, %12
store i32 %13, ptr %7, align 4
br label %B1
B1:
%14 = load i32, ptr %5, align 4
%15 = load i32, ptr %7, align 4
%16 = icmp sle i32 %14, %15
br i1 %16, label %B2, label %B3
B2:
%17 = load i32, ptr %7, align 4
%18 = load i32, ptr %5, align 4
%19 = sub nsw i32 %17, %18
store i32 %19, ptr %7, align 4
%20 = load i32, ptr %6, align 4
%21 = add nsw i32 %20, 1
store i32 %21, ptr %6, align 4
%22 = load i32, ptr %7, align 4
%23 = add nsw i32 %22, 1
store i32 %23, ptr %7, align 4
br label %B1
B3:
%24 = load i32, ptr %6, align 4
call void @llvm.lifetime.end.p0(i64 4, ptr %7) #2
call void @llvm.lifetime.end.p0(i64 4, ptr %6) #2
ret i32 %24
}
\end{lstlisting}

\textbf{LLVM \texttt{-O3} Optimization.} LLVM \texttt{-O3} performs a series of standard IR-level optimizations on the original program. First, stack-allocated scalars are promoted to SSA form via mem2reg, eliminating redundant memory operations and lifetime intrinsics. Loop-invariant computations, such as the initial division \texttt{q=n/c}, are hoisted out of the loop, and loop-carried values are represented using PHI nodes. As a result, the optimized IR produced by LLVM \texttt{-O3} expresses the wrapper-exchange process more compactly in SSA form and removes unnecessary memory traffic. However, the core computation remains a loop that repeatedly updates the number of wrappers and consumed chocolates.

\begin{lstlisting}[
frame=single,
rulecolor=\color{black},
backgroundcolor=\color{gray!10},
basicstyle=\ttfamily\small,
breaklines=true,
breakindent=0pt,
breakautoindent=false,
showstringspaces=false
]
define dso_local i32 @chocolateFeast(i32 noundef %0, i32 noundef %1, i32 noundef %2) local_unnamed_addr #0 {
B0:
%3 = sdiv i32 %0, %1
%4 = icmp slt i32 %3, %2
br i1 %4, label %B2, label %B1
B1:
%5 = phi i32 [ %9, %B1 ], [ %3, %B0 ]
%6 = phi i32 [ %7, %B1 ], [ %3, %B0 ]
%7 = add nsw i32 %6, 1
%8 = sub i32 %5, %2
%9 = add i32 %8, 1
%10 = icmp slt i32 %9, %2
br i1 %10, label %B2, label %B1
B2:
%11 = phi i32 [ %3, %B0 ], [ %7, %B1 ]
ret i32 %11
}
\end{lstlisting}

\textbf{\system.} In contrast, \system goes beyond structural IR cleanup and performs algorithm-level reasoning. After applying the same foundational optimizations as LLVM \texttt{-O3} -- such as SSA promotion, loop canonicalization, and invariant hoisting -- \system identifies a key mathematical property of the wrapper-exchange process.

Specifically, when \texttt{m} $>1$ and \texttt{q} $>0$, the number of additional chocolates obtained through wrapper exchanges can be computed in closed form as:
$$\left \lfloor \frac{q-1}{m-1}  \right \rfloor $$
Based on this observation, \system introduces a fast path that directly computes the total number of chocolates as:
$$q + \left \lfloor \frac{q-1}{m-1}  \right \rfloor $$
completely eliminating the loop in this common case. Importantly, boundary checks are preserved: the fast path is only taken when \texttt{m} $>1$, avoiding division-by-zero and ensuring semantic equivalence with the original program. For remaining cases, \system falls back to a canonicalized loop that mirrors the original behavior, including non-termination when \texttt{m} $\leq 1$.

The refined strategy generated by \system is as follows:

\begin{lstlisting}[style=mylisting]
- Promote all stack-allocated scalars to SSA (mem2reg)
  - Remove allocas (%3, %4, %5, %6, %7), loads/stores, and lifetime intrinsics.
  - Use incoming arguments directly and represent loop-carried values with PHI nodes.

- Hoist loop-invariant computations and eliminate redundant loads
  - Compute the initial quotient once (%q = sdiv i32 %n, %c).
  - Use SSA values for the loop compare and updates instead of reloading memory.

- Canonicalize the loop and add a preheader guard
  - Build a single loop header with PHIs for the running counts (consumed and wrappers).
  - Check in a preheader whether exchanges are possible (m <= wrappers). If not, return the initial quotient immediately.

- Replace memory-based updates with SSA arithmetic and preserve flags
  - In the loop body, compute wrappers = wrappers - m + 1 and consumed = consumed + 1 using nsw where appropriate, matching the original semantics.

- (*@\hl{Add a fast closed-form path for m > 1}@*)
  - (*@\hltext{When m > 1 and the initial quotient q > 0, the number of exchanges is floor((q - 1) / (m - 1)). Return q + floor((q - 1) / (m - 1)) without entering the loop.}@*)
  - (*@\hltext{Keep boundary checks: do not compute (q - 1) / (m - 1) unless m > 1 to avoid division by zero. For other cases, fall back to the loop, preserving original behavior (including non-termination for m <= 1).}@*)

- Tighten function attributes
  - The optimized function performs only pure arithmetic and has no memory side effects. Mark it nounwind, willreturn, and memory(none). Keep target-specific attributes from the original.

- Preserve loop metadata
  - Attach !llvm.loop.mustprogress to the loop latch to aid further optimizations.
\end{lstlisting}

The optimized IR generated by \system is as follows:

\begin{lstlisting}[
frame=single,
rulecolor=\color{black},
backgroundcolor=\color{gray!10},
basicstyle=\ttfamily\small,
breaklines=true,
breakindent=0pt,
breakautoindent=false,
showstringspaces=false
]
define dso_local i32 @chocolateFeast(i32 noundef %n, i32 noundef %c, i32 noundef %m) #0 {
entry:
  %q = sdiv i32 %n, %c
  %m_gt1 = icmp sgt i32 %m, 1
  %q_gt0 = icmp sgt i32 %q, 0
  %fast = and i1 %m_gt1, %q_gt0
  br i1 %fast, label %fastpath, label %preheader

fastpath:
  %q_minus1 = add nsw i32 %q, -1
  %m_minus1 = add nsw i32 %m, -1
  %extra = sdiv i32 %q_minus1, %m_minus1
  %total = add nsw i32 %q, %extra
  ret i32 %total

preheader:
  %can_loop = icmp sle i32 %m, %q
  br i1 %can_loop, label %loop, label %exit

loop:
  %wrappers = phi i32 [ %q, %preheader ], [ %wrappers.next, %loop ]
  %consumed = phi i32 [ %q, %preheader ], [ %consumed.next, %loop ]
  %tmp = sub nsw i32 %wrappers, %m
  %wrappers.next = add nsw i32 %tmp, 1
  %consumed.next = add nsw i32 %consumed, 1
  %cond = icmp sle i32 %m, %wrappers.next
  br i1 %cond, label %loop, label %exit, !llvm.loop !9

exit:
  %result = phi i32 [ %q, %preheader ], [ %consumed.next, %loop ]
  ret i32 %result
}
\end{lstlisting}

\textbf{Discussion.} This example highlights a fundamental distinction between traditional compiler optimizations and intent-driven optimization. LLVM \texttt{-O3} improves performance by restructuring the program within the same algorithmic framework, whereas \system recognizes and exploits a higher-level mathematical invariant of the algorithm itself. By replacing an iterative process with a closed-form computation under well-defined conditions, \system achieves an optimization that is fundamentally algorithmic rather than purely syntactic or structural.

\subsection{Global-coordinated Optimization}

\subsubsection{reverse\_bit}
\textbf{Program Description.} This function implements a data-dependent bit-reversal routine. Starting from an input integer \texttt{n}, it iteratively constructs an output accumulator by (1) shifting the accumulator left by one bit, (2) appending the current least-significant bit of \texttt{n}, and (3) shifting \texttt{n} right by one bit. The loop terminates when \texttt{n=0}. Intuitively, the function returns the bit-reversal of the significant prefix of \texttt{n}: it reverses the bits of \texttt{n} up to its most-significant 1-bit, while ignoring leading zeros. For example, \texttt{n}=6 (\texttt{110b}) yields \texttt{011b} (=3).

\textbf{Baseline Implementation (Unoptimized).} The unoptimized IR expresses the algorithm in a literal, step-by-step manner. It allocates stack slots for the input, the accumulator, and a temporary bit, and repeatedly performs memory loads/stores inside the loop.
\begin{lstlisting}[
frame=single,
rulecolor=\color{black},
backgroundcolor=\color{gray!10},
basicstyle=\ttfamily\small,
breaklines=true,
breakindent=0pt,
breakautoindent=false,
showstringspaces=false
]
define dso_local noundef i32 @_Z12reverse_bitsi(i32 noundef %0) #0 {
B0:
%1 = alloca i32, align 4
%2 = alloca i32, align 4
%3 = alloca i32, align 4
store i32 %0, ptr %1, align 4
call void @llvm.lifetime.start.p0(i64 4, ptr %2) #2
store i32 0, ptr %2, align 4
call void @llvm.lifetime.start.p0(i64 4, ptr %3) #2
br label %B1
B1:
%4 = load i32, ptr %1, align 4
%5 = icmp ne i32 %4, 0
br i1 %5, label %B2, label %B3
B2:
%6 = load i32, ptr %2, align 4
%7 = shl i32 %6, 1
store i32 %7, ptr %2, align 4
%8 = load i32, ptr %1, align 4
%9 = and i32 %8, 1
store i32 %9, ptr %3, align 4
%10 = load i32, ptr %3, align 4
%11 = load i32, ptr %2, align 4
%12 = or i32 %10, %11
store i32 %12, ptr %2, align 4
%13 = load i32, ptr %1, align 4
%14 = ashr i32 %13, 1
store i32 %14, ptr %1, align 4
br label %B1
B3:
%15 = load i32, ptr %2, align 4
call void @llvm.lifetime.end.p0(i64 4, ptr %3) #2
call void @llvm.lifetime.end.p0(i64 4, ptr %2) #2
ret i32 %15
}
\end{lstlisting}

\textbf{LLVM \texttt{-O3} Optimization.} LLVM \texttt{-O3} performs canonical IR cleanups and loop simplification. It promotes stack-allocated variables to SSA (mem2reg), removes lifetime intrinsics, and represents the loop-carried state using PHI nodes: one PHI for the evolving input value and one for the accumulator.
\begin{lstlisting}[
frame=single,
rulecolor=\color{black},
backgroundcolor=\color{gray!10},
basicstyle=\ttfamily\small,
breaklines=true,
breakindent=0pt,
breakautoindent=false,
showstringspaces=false
]
define dso_local noundef i32 @_Z12reverse_bitsi_opt(i32 noundef %0) local_unnamed_addr #0 {
B0:
%1 = icmp eq i32 %0, 0
br i1 %1, label %B2, label %B1
B1:
%2 = phi i32 [ %7, %B1 ], [ %0, %B0 ]
%3 = phi i32 [ %6, %B1 ], [ 0, %B0 ]
%4 = shl i32 %3, 1
%5 = and i32 %2, 1
%6 = or disjoint i32 %5, %4
%7 = ashr i32 %2, 1
%8 = icmp ult i32 %2, 2
br i1 %8, label %B2, label %B1
B2:
%9 = phi i32 [ 0, %B0 ], [ %6, %B1 ]
ret i32 %9
}
\end{lstlisting}

\textbf{\system.} \system identifies the loop as an instance of bit reversal and realizes it using a coordinated transformation that spans multiple abstraction levels: (1) semantic recognition of the loop’s algorithmic intent, (2) selection of hardware-mappable LLVM intrinsics, and (3) careful boundary handling to preserve defined behavior. Concretely, \system replaces the entire loop with two compiler intrinsics: \texttt{llvm.bitreverse.i32} to reverse all 32 bits of \texttt{n}, and \texttt{llvm.ctlz.i32} (with \texttt{is\_zero\_undef=false}) to compute the number of leading zeros.

The refined strategy generated by \system is as follows:

\begin{lstlisting}[style=mylisting]
- (*@\hl{Replace the loop with two data-dependent instructions that the backend can map to hardware:}@*) (*@\hltext{llvm.bitreverse.i32 to reverse all 32 bits, and llvm.ctlz.i32 (with is\_zero\_undef=false) to compute the number of leading zeros. The final result is lshr(bitreverse(n), ctlz(n)), which exactly matches the original loop’s semantics for non-negative inputs and avoids undefined behavior by guarding n==0.}@*)
- Keep a boundary check for n==0 to avoid shifting by the full width (which is undefined). This preserves correctness while enabling a branch-free fast path for all non-zero inputs.
- Avoid allocas, lifetime intrinsics, and redundant loads/stores. Operate in SSA form and mark the function readnone, nounwind, and willreturn to let LLVM perform more aggressive optimizations and inlining if beneficial.
- Note: The original IR uses an arithmetic right shift (ashr) on a signed int and a while(n != 0) termination condition, which leads to non-termination for negative inputs. If negative inputs are possible, you should constrain callers to pass non-negative values or switch the original algorithm to logical shifts. The optimized version below preserves the zero boundary check and provides well-defined behavior for all inputs, returning 0 for n==0 and a fast reverse for n>0.
\end{lstlisting}

The optimized IR generated by \system is as follows:
\begin{lstlisting}[
frame=single,
rulecolor=\color{black},
backgroundcolor=\color{gray!10},
basicstyle=\ttfamily\small,
breaklines=true,
breakindent=0pt,
breakautoindent=false,
showstringspaces=false
]
define dso_local noundef i32 @_Z12reverse_bitsi(i32 noundef %n) local_unnamed_addr #0 {
entry:
  %iszero = icmp eq i32 %n, 0
  br i1 %iszero, label %ret, label %fast

fast:
  %brev = call i32 @llvm.bitreverse.i32(i32 %n)
  %clz = call i32 @llvm.ctlz.i32(i32 %n, i1 false)
  %res = lshr i32 %brev, %clz
  br label %ret

ret:
  %r = phi i32 [ 0, %entry ], [ %res, %fast ]
  ret i32 %r
}
\end{lstlisting}

\subsubsection{\_Z3conii}

\textbf{Program Description.} The function  returns a Boolean-like integer (\(0\) or \(1\)) based on a set of guarded conditions. Intuitively, it checks whether the integer \(x\) equals the number of decimal digits of \(y\), under additional validity constraints: \(x > 0\), \(y \in [1, m)\) where \(m\) is a global threshold, and a special boundary case when \(y = 0\). The unoptimized implementation computes the digit count as \(\lfloor \log_{10}(y) \rfloor + 1\) and compares it with \(x\), returning \(1\) iff they match; otherwise it returns \(0\).

\textbf{Baseline Implementation (Unoptimized).} The unoptimized IR implements the logic using multiple basic blocks and stack-based temporaries. It computes \(\lfloor \log_{10}(y) \rfloor + 1\) via integer-to-floating-point conversion, a math-library call to \texttt{log10}, and a floating-point-to-integer cast.

\begin{lstlisting}[
frame=single,
rulecolor=\color{black},
backgroundcolor=\color{gray!10},
basicstyle=\ttfamily\small,
breaklines=true,
breakindent=0pt,
breakautoindent=false,
showstringspaces=false
]
define dso_local noundef i32 @_Z3conii(i32 noundef %0, i32 noundef %1) #0 {
B0:
%2 = alloca i32, align 4
%3 = alloca i32, align 4
%4 = alloca i32, align 4
store i32 %0, ptr %3, align 4
store i32 %1, ptr %4, align 4
%5 = load i32, ptr %3, align 4
%6 = icmp sle i32 %5, 0
br i1 %6, label %B1, label %B2
B1:
store i32 0, ptr %2, align 4
br label %B10
B2:
%7 = load i32, ptr %4, align 4
%8 = icmp eq i32 %7, 0
br i1 %8, label %B3, label %B5
B3:
%9 = load i32, ptr %3, align 4
%10 = icmp eq i32 %9, 1
br i1 %10, label %B4, label %B5
B4:
store i32 1, ptr %2, align 4
br label %B10
B5:
%11 = load i32, ptr %4, align 4
%12 = icmp sle i32 1, %11
br i1 %12, label %B6, label %B9
B6:
%13 = load i32, ptr %4, align 4
%14 = load i32, ptr @m, align 4
%15 = icmp slt i32 %13, %14
br i1 %15, label %B7, label %B9
B7:
%16 = load i32, ptr %3, align 4
%17 = load i32, ptr %4, align 4
%18 = sitofp i32 %17 to double
%19 = call double @log10(double noundef %18) #2
%20 = fptosi double %19 to i32
%21 = add nsw i32 %20, 1
%22 = icmp eq i32 %16, %21
br i1 %22, label %B8, label %B9
B8:
store i32 1, ptr %2, align 4
br label %B10
B9:
store i32 0, ptr %2, align 4
br label %B10
B10:
%23 = load i32, ptr %2, align 4
ret i32 %23
}
\end{lstlisting}

\textbf{LLVM \texttt{-O3} Optimization.}
LLVM \texttt{-O3} applies standard IR-level optimizations, including promoting stack variables to SSA form, collapsing redundant basic blocks, and simplifying control flow using PHI nodes and \texttt{select} instructions. These transformations reduce memory traffic and streamline the control flow. However, the optimized IR still relies on the floating-point digit-count computation using \texttt{log10}. 

\begin{lstlisting}[
frame=single,
rulecolor=\color{black},
backgroundcolor=\color{gray!10},
basicstyle=\ttfamily\small,
breaklines=true,
breakindent=0pt,
breakautoindent=false,
showstringspaces=false
]
define dso_local noundef range(i32 0, 2) i32 @_Z3conii_opt(i32 noundef %0, i32 noundef %1) local_unnamed_addr #0 {
B0:
%2 = icmp slt i32 %0, 1
br i1 %2, label %B5, label %B1
B1:
%3 = icmp eq i32 %1, 0
%4 = icmp eq i32 %0, 1
%5 = and i1 %4, %3
br i1 %5, label %B5, label %B2
B2:
%6 = icmp sgt i32 %1, 0
%7 = load i32, ptr @m, align 4
%8 = icmp sgt i32 %7, %1
%9 = select i1 %6, i1 %8, i1 false
br i1 %9, label %B3, label %B4
B3:
%10 = uitofp nneg i32 %1 to double
%11 = tail call double @log10(double noundef %10) #2
%12 = fptosi double %11 to i32
%13 = add nsw i32 %12, 1
%14 = icmp eq i32 %13, %0
br i1 %14, label %B5, label %B4
B4:
br label %B5
B5:
%15 = phi i32 [ 0, %B4 ], [ 0, %B0 ], [ 1, %B1 ], [ 1, %B3 ]
ret i32 %15
}
\end{lstlisting}

\textbf{\system.}
\system performs a global-coordinated optimization by combining data-flow analysis with semantic reasoning over library calls. By tracing the provenance of the argument passed to \texttt{log10}, \system recognizes that the computation operates on the integer input \(y\) and that its numeric domain is tightly constrained by preceding control-flow guards. This enables \system to reinterpret the floating-point expression \(\lfloor \log_{10}(y) \rfloor + 1\) as an integer digit-count operation rather than a general library call.

Based on this reinterpretation, \system replaces the floating-point conversion and library call with a semantically equivalent integer formulation. The digit count is computed exactly using a sequence of comparisons against constant powers of ten (e.g., \(y \ge 10, 100, 1000, \ldots\)), followed by accumulation of the Boolean results.

The refined strategy generated by \system is as follows:

\begin{lstlisting}[style=mylisting]
- Promote stack slots to SSA: Remove alloca/store/load of the input arguments and the result. Use the incoming arguments %0 and %1 directly and produce the result via straight SSA, returning directly from blocks.
- Simplify the entry guards: Fold the initial "%0 <= 0" check into a single branch that returns 0 immediately. This avoids extra blocks and memory traffic.
- Merge the two-step zero check: Replace the "%1 == 0" then "%0 == 1" chain with a single block returning zext(%0 == 1) when %1 == 0. This keeps the necessary boundary check for log10(0), but is branch-minimal.
- Canonicalize the range check: Compute "inrange = (%1 >= 1) && (%1 < m)" as a single predicate and branch only once based on it. This preserves the boundary checks while cutting down on control flow.
- (*@\hl{Strength-reduce and eliminate the floating-point path:}@*) (*@\hltext{Replace the floor(log10(\%1))  + 1 digit-count using the math library with an exact, branchless integer computation of decimal digits based on comparisons against constant powers of 10. This avoids FP conversion, the library call, and rounding concerns, and is typically much faster and more predictable than -O3's FP-based codegen.}@*)
- Reduce exits: Return directly from each decision point (early returns) rather than storing to a temporary and reading it back or building a complex control flow with PHIs. This enables simpler code generation and tail duplication when profitable.
- Tighten attributes: Mark the function as readonly (it only reads m and does not write memory). Keep nounwind/mustprogress, which allows further optimization and better code generation without changing semantics.
\end{lstlisting}

The optimized IR generated by \system is as follows:

\begin{lstlisting}[
frame=single,
rulecolor=\color{black},
backgroundcolor=\color{gray!10},
basicstyle=\ttfamily\small,
breaklines=true,
breakindent=0pt,
breakautoindent=false,
showstringspaces=false
]
define dso_local noundef i32 @_Z3conii(i32 noundef %x, i32 noundef %y) #0 {
entry:
  %cmp_x_le_0 = icmp sle i32 %x, 0
  br i1 %cmp_x_le_0, label %ret0, label %check_y_zero

ret0:
  ret i32 0

check_y_zero:
  %y_is_zero = icmp eq i32 %y, 0
  br i1 %y_is_zero, label %y_zero_case, label %inrange_check

y_zero_case:
  %x_is_one = icmp eq i32 %x, 1
  %ret_y0 = zext i1 %x_is_one to i32
  ret i32 %ret_y0

inrange_check:
  %ge1 = icmp sge i32 %y, 1
  %mval = load i32, ptr @m, align 4, !tbaa !5
  %ltm = icmp slt i32 %y, %mval
  %inrange = and i1 %ge1, %ltm
  br i1 %inrange, label %digits_compute, label %ret0b

ret0b:
  ret i32 0

digits_compute:
  %c1  = icmp sge i32 %y, 10
  %c2  = icmp sge i32 %y, 100
  %c3  = icmp sge i32 %y, 1000
  %c4  = icmp sge i32 %y, 10000
  %c5  = icmp sge i32 %y, 100000
  %c6  = icmp sge i32 %y, 1000000
  %c7  = icmp sge i32 %y, 10000000
  %c8  = icmp sge i32 %y, 100000000
  %c9  = icmp sge i32 %y, 1000000000

  %t1  = zext i1 %c1 to i32
  %t2  = zext i1 %c2 to i32
  %t3  = zext i1 %c3 to i32
  %t4  = zext i1 %c4 to i32
  %t5  = zext i1 %c5 to i32
  %t6  = zext i1 %c6 to i32
  %t7  = zext i1 %c7 to i32
  %t8  = zext i1 %c8 to i32
  %t9  = zext i1 %c9 to i32

  %s1  = add nuw nsw i32 %t1, %t2
  %s2  = add nuw nsw i32 %t3, %t4
  %s3  = add nuw nsw i32 %t5, %t6
  %s4  = add nuw nsw i32 %t7, %t8
  %s5  = add nuw nsw i32 %s1, %s2
  %s6  = add nuw nsw i32 %s3, %s4
  %s7  = add nuw nsw i32 %s5, %s6
  %s8  = add nuw nsw i32 %s7, %t9
  %digits = add nuw nsw i32 %s8, 1

  %eqdigits = icmp eq i32 %x, %digits
  %retv = zext i1 %eqdigits to i32
  ret i32 %retv
}
\end{lstlisting}

\paragraph{Discussion.}
This example highlights the advantage of \system’s global-coordinated optimization guided by optimization intent. \system recognizing the semantic role of the computation, validating its applicability via global guards, and selecting an implementation that better aligns with the target execution model. As a result, \system performs an optimization that spans multiple abstraction layers, demonstrating how global coordination enables performance improvements beyond the reach of conventional compiler pipelines.


\end{document}